\documentclass{article}

\usepackage{arxiv}

\usepackage[utf8]{inputenc} 
\usepackage[T1]{fontenc}    
\usepackage{hyperref}       
\usepackage{url}            
\usepackage{booktabs}       
\usepackage{amsfonts}       
\usepackage{nicefrac}       
\usepackage{microtype}      
\usepackage{lipsum}		
\usepackage{setspace}
\usepackage{graphicx}
\usepackage{doi}
\usepackage{amsmath}
\usepackage[numbers,sort&compress]{natbib}

\title{Quadrene: A Novel Quasi-2D Carbon Allotrope with High Carrier Mobility}

\author{%
\parbox{0.95\linewidth}{\centering
\href{https://orcid.org/0000-0003-4699-5886}{\includegraphics[scale=0.09]{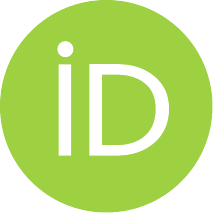}\hspace{1mm}}Kleuton A.~L.~Lima\textsuperscript{1},
\href{https://orcid.org/0000-0002-8366-7227}{\includegraphics[scale=0.09]{icons/orcid.pdf}\hspace{1mm}}Jos\'e A.~dos S.~Laranjeira\textsuperscript{2},
\href{https://orcid.org/0009-0003-1216-1629}{\includegraphics[scale=0.09]{icons/orcid.pdf}\hspace{1mm}}Neymar J.~N.~Cavalcante\textsuperscript{3},
\href{https://orcid.org/0000-0001-7653-0428}{\includegraphics[scale=0.09]{icons/orcid.pdf}\hspace{1mm}}Nicolas F.~Martins\textsuperscript{2},
\href{https://orcid.org/0000-0002-5217-7145}{\includegraphics[scale=0.09]{icons/orcid.pdf}\hspace{1mm}}Julio R. Sambrano\textsuperscript{2},
\href{https://orcid.org/0000-0003-0145-8358}{\includegraphics[scale=0.09]{icons/orcid.pdf}\hspace{1mm}}Douglas S.~Galv\~ao\textsuperscript{4},
and
\href{https://orcid.org/0000-0001-7468-2946}{\includegraphics[scale=0.09]{icons/orcid.pdf}\hspace{1mm}}Luiz A.~Ribeiro Jr\textsuperscript{2,$\dag$} \\
\vspace{0.6em}
{\normalfont\normalsize
\textsuperscript{1}Department of Applied Physics and Center for Computational Engineering and Sciences, State University of Campinas, Campinas, 13083-859, SP, Brazil\\
\textsuperscript{2}Modeling and Molecular Simulation Group, S\~ao Paulo State University (UNESP), School of Sciences, Bauru 17033-360, SP, Brazil\\
\textsuperscript{3}State University of Piauí, Campus Prof. Antônio Giovanni Alves de Sousa, Av. Mal. Castelo Branco, 64260-000, Piripiri, PI, Brazil\\
\textsuperscript{4}Computational Materials Laboratory, LCCMat, Institute of Physics, University of Bras\'ilia, 70910-900, Bras\'ilia, Federal District, Brazil\\
\vspace{0.6em}
\href{https://scholar.google.com.br/citations?user=e_2ul00AAAAJ&hl=pt-BR}{\includegraphics[scale=0.05]{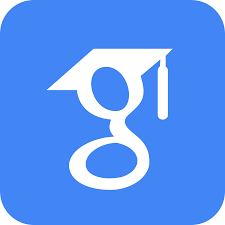}} \href{https://www.linkedin.com/in/kleuton-antunes-1023b4392/}{\includegraphics[scale=0.05]{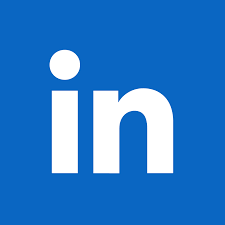}}\hspace{0.1cm}\texttt{\textsuperscript{*}kleuton@unicamp.br} \\
\vspace{0.1cm}
\href{https://scholar.google.com/citations?user=EgsxcaUAAAAJ\&hl=pt-BR}{\includegraphics[scale=0.05]{icons/gscholar.png}} \href{www.linkedin.com/in/luiz-ribeiro-164221225}{\includegraphics[scale=0.05]{icons/linkedin.png}}\hspace{0.1cm}\texttt{\textsuperscript{$\dag$}ribeirojr@unb.br}\\
}
}%
}

\renewcommand{\shorttitle}{Kleuton A.~L.~Lima et al.}

\hypersetup{
pdftitle={kagome-goldene},
pdfsubject={kagome, goldene},
pdfauthor={Carlos M. O. Bastos, Luiz A. Ribeiro Jr},
pdfkeywords={Two-dimensional metals, Kagome lattice, Goldene, Lattice dynamics, Strain engineering, First-principles calculations.},
}

\begin{document}
\maketitle

\onehalfspacing

\begin{abstract}
We present a comprehensive first-principles investigation of a novel carbon allotrope characterized by quasi-tetragonal atomic motifs and quasi-two-dimensional structural behavior. Structural analysis reveals an open framework composed of alternating diamond-like and square units, while thermodynamic assessments indicate a negative formation energy, suggesting high intrinsic stability. Phonon spectra confirm dynamical robustness, and \textit{ab initio} molecular dynamics simulations at 1000~K validate its thermal resilience. Furthermore, the system exhibits an indirect bandgap of 1.58 eV at the HSE06 level, anisotropic mechanical behavior, and a broadband optical response, reinforcing its potential for nanoelectronic and optoelectronic applications. The highly anisotropic mechanical behavior is characterized by an in-plane Young's modulus ranging from 80 to 550~GPa depending on crystallographic direction. Additionally, the electronic transport properties exhibit pronounced anisotropy, with hole mobilities reaching up to $5.83 \times 10^{6}$ cm$^{2}$/V·s and electron mobilities up to $6.40 \times 10^{6}$ cm$^{2}$/V·s along different crystallographic directions, highlighting the material's promise for directionally selective nanoelectronic device applications.
\end{abstract}


\section{Introduction}

The discovery of graphene ushered in a new era in materials science, revealing that two-dimensional (2D) carbon materials can exhibit extraordinary physical properties such as high mechanical strength, exceptional electrical conductivity, and tunable optical responses~\cite{geim2009graphene,geim2007rise}. Since then, various 2D carbon allotropes have been proposed and synthesized \cite{hirsch2010era}, including graphyne \cite{desyatkin2022scalable,aliev2025planar}, graphdiyne \cite{li2010architecture}, monolayer fullerene network \cite{hou2022synthesis,meirzadeh2023few}, and biphenylene network~\cite{fan2021biphenylene}, each expanding the structural diversity and functional potential of carbon-based nanomaterials. These new architectures are often characterized by the inclusion of multiple ring sizes, acetylenic linkages, or non-hexagonal symmetries, offering a broader design space for engineering electronic, mechanical, and optical characteristics.

A particularly promising strategy in carbon materials design lies in the construction of frameworks incorporating non-traditional polygonal units, such as squares, octagons, and dodecagons, arranged in open networks \cite{jana2021emerging,xu2013graphene,tiwari2016magical,enyashin2011graphene,lima2025structural,lima2025petal,tromer2023mechanical,junior2023irida,lima2025anthraphenylenes,lima20252Ddodeca,LIMA2024109455,bafekry2024layered,bafekry2021biphenylene,bafekry2022tunable}. These configurations can give rise to novel topologies, directional bonding, and anisotropic behavior, potentially leading to applications in nanoelectronics, energy storage, and photonics~\cite{zhang20162d,yamashita2019nonlinear,burghard2009carbon}. Among them, recent advances in computational predictions have identified carbon allotropes with quasi-tetragonal symmetry and low dimensionality that combine planar features with distinct out-of-plane buckling or hierarchical pore organization~\cite{xing2023physical}.

In this work, we introduce a novel quasi-two-dimensional carbon allotrope, named Quadrene. Through a combination of first-principles calculations and detailed analysis of electronic, dynamical, and mechanical properties, we demonstrate that this material is both thermally and dynamically stable. Unlike graphene, the proposed allotrope exhibits a small indirect bandgap of 0.80 eV, a key feature for semiconducting behavior in nanoelectronic platforms. Furthermore, our results reveal that the material retains good mechanical strength, with calculated in-plane Young's moduli reaching 550 GPa. These properties highlight the promising potential of this carbon framework for applications in flexible semiconducting devices and strain-tunable electronics.

\section{Methodology}

All calculations were performed using first-principles density functional theory (DFT), as implemented in the CASTEP software package~\cite{clark2005first}. The exchange-correlation interactions were treated within the generalized gradient approximation (GGA) using the Perdew–Burke–Ernzerhof (PBE) functional~\cite{perdew1996generalized}, in combination with norm-conserving pseudopotentials. Geometry optimizations were performed using the Broyden–Fletcher–Goldfarb–Shanno (BFGS) minimization algorithm~\cite{head1985broyden,PFROMMER1997233}, ensuring high accuracy through convergence thresholds of $10^{-5}$~eV for total energy, $10^{-3}$~eV/\r{A} for atomic forces, and $10^{-2}$~GPa for the residual stress.

To prevent artificial interactions between periodic images along the non-periodic axis, a vacuum region of 20~\r{A} was introduced. The Brillouin zone was sampled using a Monkhorst–Pack grid of $5\times5\times1$ for structural relaxations, while a denser $20\times20\times1$ mesh was adopted for electronic structure and optical property analyses. The complex dielectric function, from which absorption and reflectivity spectra were derived, was computed following the formalism outlined in Ref.~\cite{lima2023dft}.

Phonon dispersion relations were obtained through density functional perturbation theory (DFPT), employing a $5\times5\times1$ q-point grid, a finite displacement amplitude of 0.05~\r{A}$^{-1}$, and a force convergence threshold of $10^{-5}$~eV/\r{A}$^{2}$. To assess thermal resilience, \textit{ab initio} molecular dynamics (AIMD) simulations were carried out at 1000~K within the NVT ensemble using a Nose–Hoover thermostat.

The elastic response was characterized by applying small deformations to the relaxed structure and evaluating the resulting stress–strain relationships. Elastic constants were extracted and used to compute the in-plane Young’s moduli and Poisson’s ratios~\cite{Zuo:gl0256,10.1063/1.1709944}.

\section{Results}

\begin{figure}[ht]
    \centering
    \includegraphics[width=\linewidth]{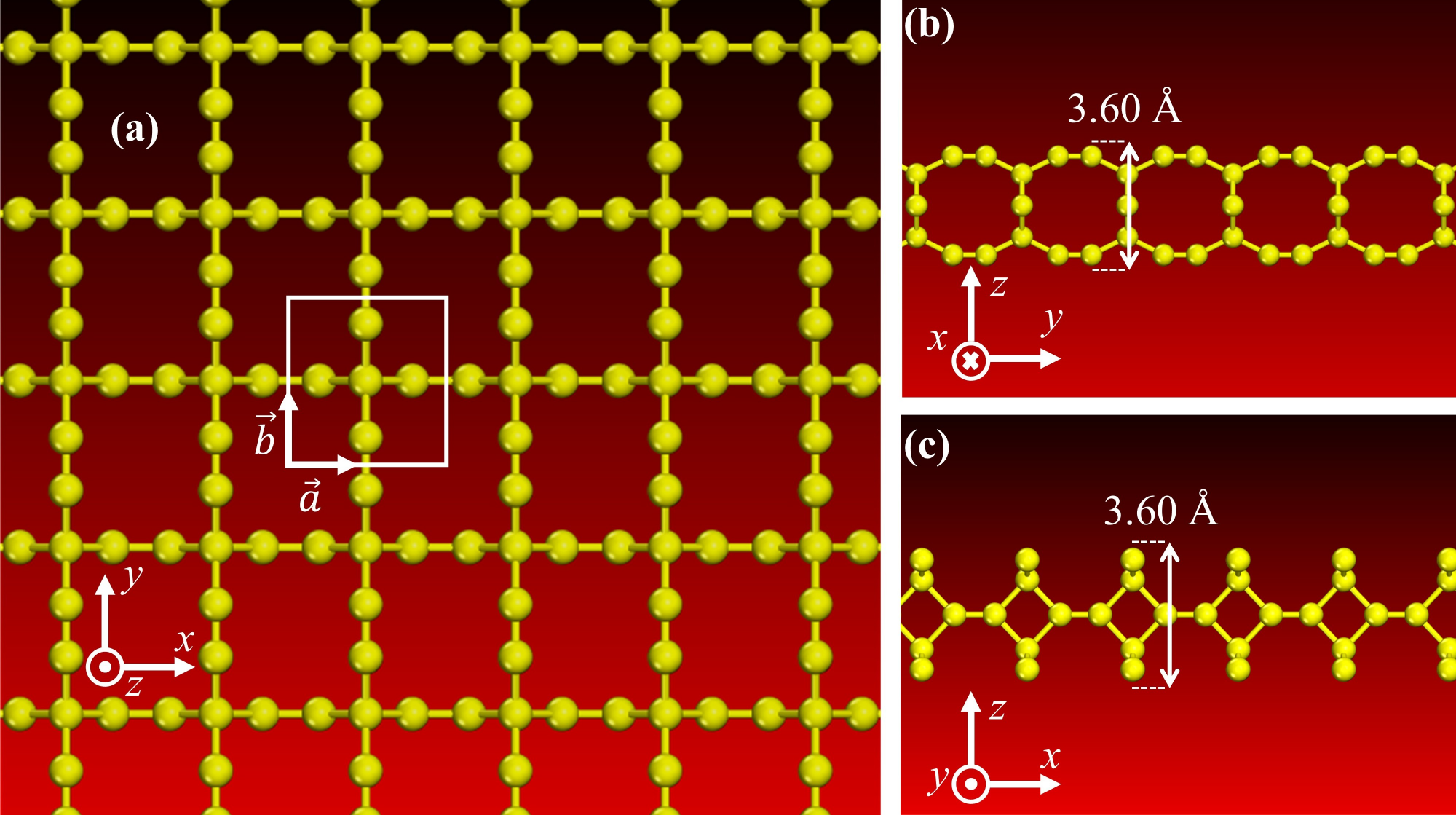}
    \caption{Atomic structure of the proposed quasi-2D carbon material. (a) Top view showing the rectangular unit cell defined by vectors $\vec{a}$ and $\vec{b}$. (b) and (c) Side views along the $x$- and $y$-directions, respectively, revealing a finite thickness of 3.60~\AA{} due to the buckled configuration.}
    \label{fig:structure}
\end{figure}

Fig.~\ref{fig:structure} illustrates the optimized atomic structure of the newly proposed quasi-two-dimensional carbon allotrope, designated as Quadrene. In Fig.~\ref{fig:structure}(a), the top view reveals a planar network of carbon atoms arranged to form a regular pattern of rectangular pores, oriented along both the $\vec{a}$ and $\vec{b}$ lattice directions. The optimized lattice constants of $a = 3.46$~\AA{} and $b = 3.84$~\AA{}. This configuration maintains orthorhombic symmetry, as characterized by the \textit{Pmmm} space group (No.~47). The structure is invariant under the symmetry operations of the \textit{D}$_{2h}$ point group. 

A detailed geometric analysis reveals the simultaneous presence of $sp$, $sp^2$, and $ sp^3$-hybridized carbon atoms, each occupying distinct regions according to their bonding geometries. The $sp$-hybridized carbons are found at positions where carbon atoms form nearly linear chains, typically observed at the sides of the rectangular pores, where the bond angles approach $180^\circ$. In contrast, the $sp^2$-hybridized carbons are located perpendicularly to the monolayer plane, as evidenced in the side views of the figure (Fig. \ref{fig:structure}c). This arrangement results in a finite thickness of 3.60~\AA{}. These atoms exhibit trigonal planar coordination with bond angles close to $120^\circ$, supporting the formation of extended conjugated $\pi$-systems throughout the two-dimensional lattice. The $sp^3$-hybridized carbon atoms, on the other hand, are along the segments that define the vertices of the rectangular motifs. These atoms possess a tetrahedral coordination environment. 

\begin{figure}[ht]
\centering
\includegraphics[width=0.7\linewidth]{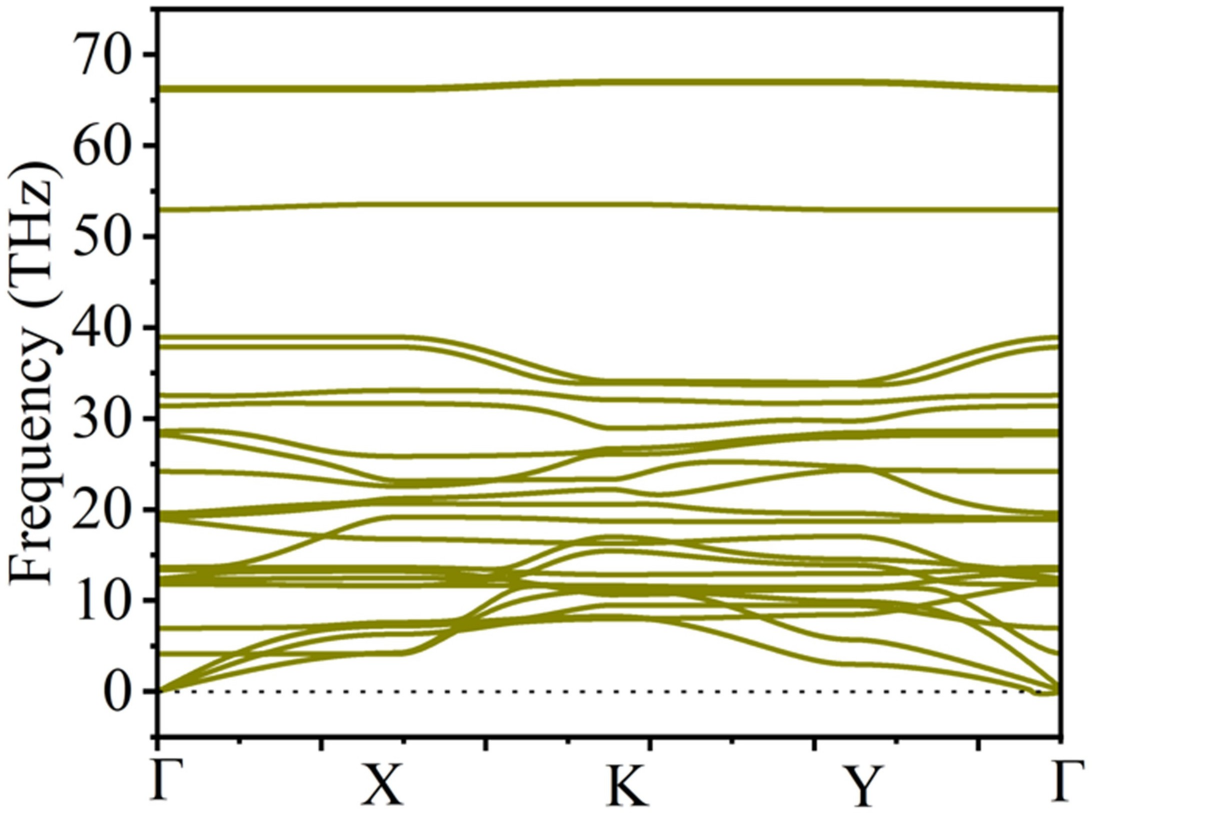}
\caption{(Phonon dispersion curves of Quadrene calculated along high-symmetry directions in the Brillouin zone.}
\label{fig:phonon}
\end{figure}

To assess the structural integrity of Quadrene, we evaluated its dynamical and thermal stability using phonon dispersion and \textit{ab initio} molecular dynamics (AIMD) simulations. Fig.~\ref{fig:phonon} shows the phonon band structure calculated along the high-symmetry path $\Gamma$–X–K–Y–$\Gamma$. The absence of imaginary modes throughout the Brillouin zone confirms the dynamic stability of the monolayer.

The phonon spectrum spans a wide frequency range, with optical branches reaching approximately 68 THz. The high-frequency modes between 60 and 70~THz originate primarily from the stretching vibrations of linear $sp$-hybridized carbon atoms within acetylene-like (C$\equiv$C) bonds~\cite{PhysRevB.92.245434}.

\begin{figure}[ht]
\centering
\includegraphics[width=0.8\linewidth]{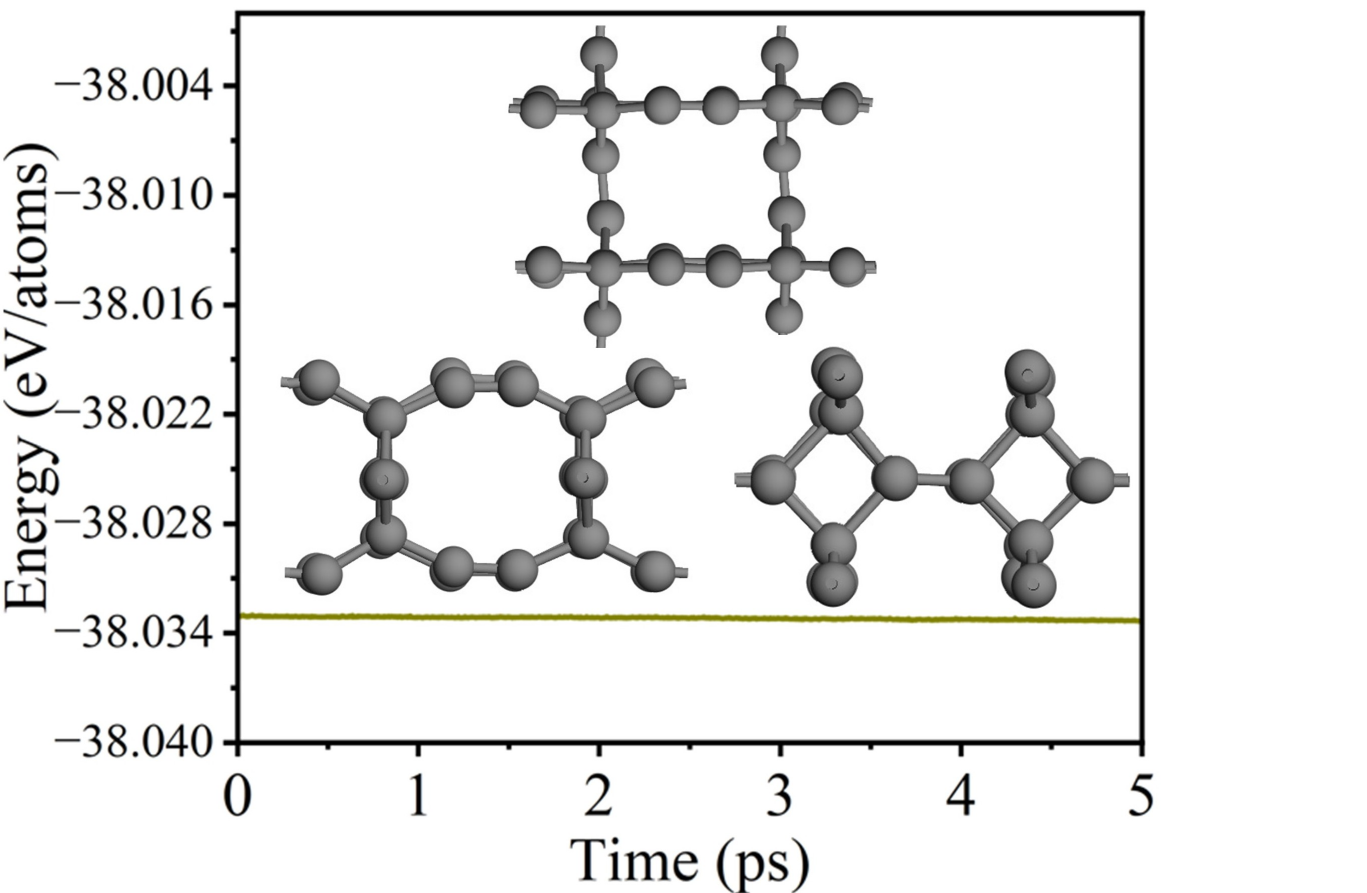}
\caption{Total energy per atom as a function of time during AIMD simulations at 1000~K, with representative atomic snapshots illustrating structural preservation throughout the simulation.}
\label{fig:aimd}
\end{figure}

Thermal stability was further examined through AIMD simulations at 1000~K in the canonical (NVT) ensemble over a 5~ps trajectory. The temporal evolution of the total energy per atom is shown in Fig.~\ref{fig:aimd}. The energy profile exhibits minimal fluctuations around a stable average of approximately -38.03 eV/atom, with no abrupt changes indicative of structural degradation. 

To evaluate the mechanical robustness and anisotropy of Quadrene, we computed its elastic constants and derived key mechanical parameters, including Poisson's ratio and Young’s modulus. The calculated in-plane elastic stiffness constants are $C_{11} = 263.27$~GPa, $C_{22} = 506.10$~GPa, $C_{12} = -18.72$~GPa, and $C_{44} = 19.13$~GPa. These values satisfy the Born–Huang mechanical stability criteria for rectangular systems, which require $C_{11}C_{22} - C_{12}^2 > 0$ and $C_{44} > 0$~\cite{PhysRevB.90.224104,doi:10.1021/acs.jpcc.9b09593}.

\begin{figure}[ht]
    \centering
    \includegraphics[width=\textwidth]{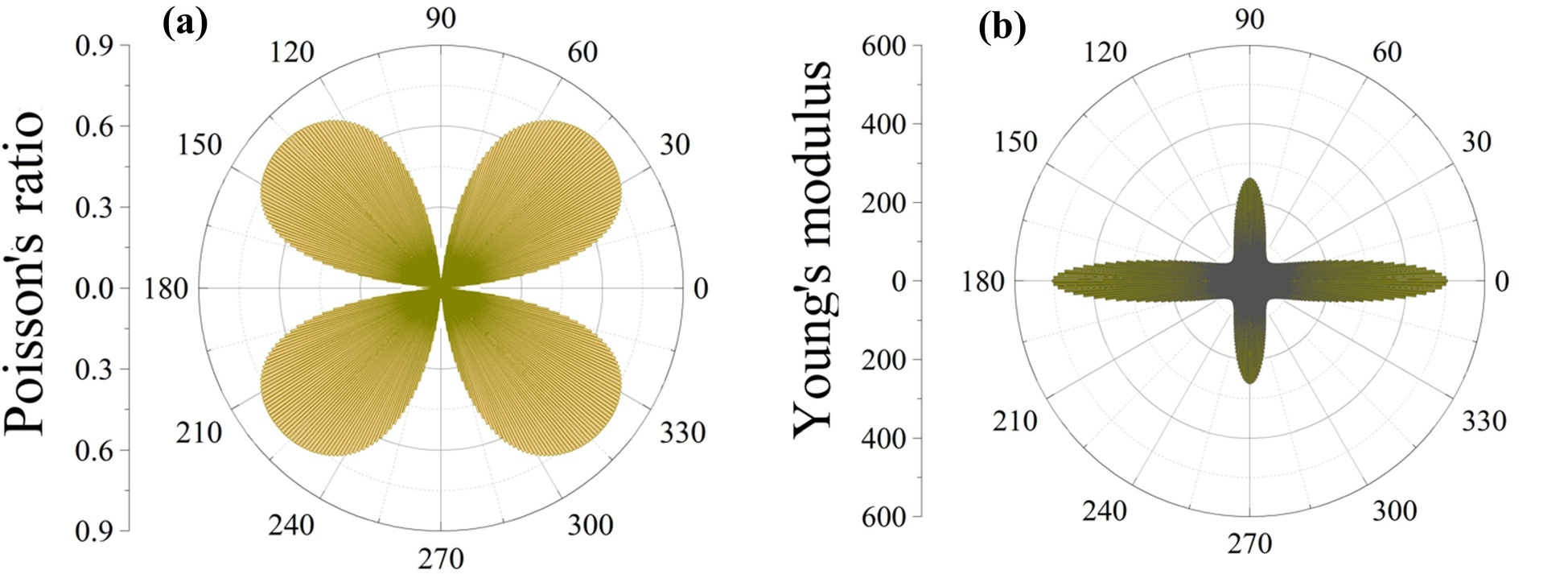}
    \caption{Mechanical response of Quadrene. (a) Polar plot of the directional Poisson's ratio, showing a four-lobed anisotropic pattern. (b) Directional dependence of the in-plane Young’s modulus, revealing marked anisotropy between the $x$ and $y$ directions.}
    \label{fig:mech}
\end{figure}

Fig.~\ref{fig:mech}(a) presents the angular dependence of Poisson's ratio $\nu(\theta)$ in polar coordinates. The four-lobed pattern reflects the quasi-tetragonal character of the structure, with $\nu$ reaching its maximum value of 0.80 along diagonal directions at $\theta = \frac{k\pi}{4} + n\pi$ ($k$ odd, $n \in \mathbb{Z}$), and vanishing along the principal symmetry axes at $\theta = n\pi/2$ ($n \in \mathbb{Z}$). This anisotropy in lateral deformation under uniaxial stress further confirms the quasi-2D nature of the lattice.

Fig.~\ref{fig:mech}(b) displays the directional Young's modulus $E(\theta)$. A pronounced mechanical anisotropy is evident, with $E(\theta)$ ranging from 259~GPa along the $x$-axis to 506~GPa along the $y$-axis. This high stiffness, particularly along the $y$-axis, stems from the alignment of carbon-carbon bonds within the octagonal-like motifs. 

\begin{figure}[ht]
    \centering
    \includegraphics[width=\textwidth]{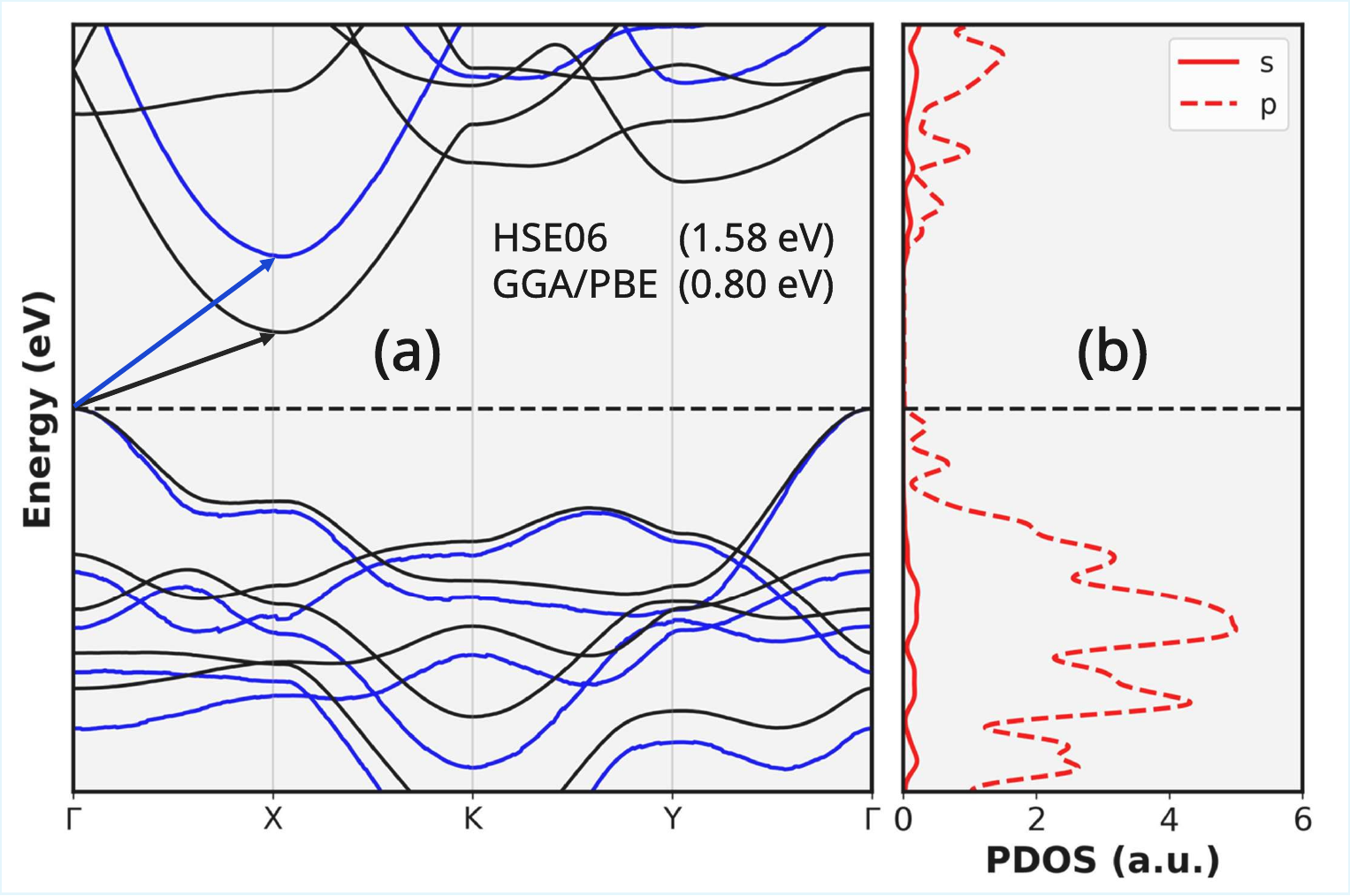}
    \caption{Electronic structure of Quadrene. (a) Electronic band structure computed using PBE and HSE06 functionals along high-symmetry directions, indicating an indirect bandgap with the valence band maximum (VBM) at the $\Gamma$ point and the conduction band minimum (CBM) at the X point. (b) Projected density of states (PDOS) showing that $p$-orbitals are the dominant contributors near the Fermi level.}
    \label{fig:elec}
\end{figure}

The electronic band structure and projected density of states (PDOS) of Quadrene are presented in Fig.~\ref{fig:elec}. As shown in Fig.~\ref{fig:elec}(a), the band structure was calculated using both the PBE and HSE06 functionals. Quadrene exhibits semiconducting behavior with an indirect bandgap of 0.80 eV at the PBE level and 1.58 eV using the HSE06 hybrid functional. In both cases, the valence band maximum (VBM) is located at the $\Gamma$ point, while the conduction band minimum (CBM) lies at X point. 

As illustrated in Fig.~\ref{fig:elec}(b), the projected density of states (PDOS) indicates that the valence band maximum (VBM) and conduction band minimum (CBM) are predominantly composed of $p$ states. At the same time, the $s$ contributions are mainly located at deeper energies (from $-2$ to $-4$~eV). This behavior reflects the coexistence of $sp$, $sp^2$, and $sp^3$ hybridizations in the structure: the $s$-dominated $\sigma$ bonds, associated with $sp^3$ orbitals, provide mechanical rigidity and structural stability, whereas the $p$-dominated $\pi$ and $\pi^{*}$ states, derived from $sp$ and $sp^2$ hybridizations, govern the transport properties. Such $\sigma$--$\pi$ mixing is responsible for the emergence of an indirect band gap of 0.80~eV and a larger direct gap of 1.58~eV.

Importantly, the presence of a finite bandgap distinguishes Quadrene from pristine graphene and other metallic-like 2D carbon allotropes~\cite{enyashin2011graphene}. This intrinsic semiconducting character overcomes one of the significant limitations of graphene, for instance, in digital logic applications, offering a promising platform for field-effect transistors (FETs), photodetectors, and other semiconductor-based nanodevices~\cite{https://doi.org/10.1002/adom.201900019, https://doi.org/10.1002/adma.201400349}. Moreover, the moderate value of the gap suggests potential for tuning via external strain~\cite{doi:10.1021/acs.chemmater.7b00453, C5RA01586C, LARANJEIRA2024114418}, doping~\cite{HUANG2024115598, WANG2024114613, MARTINS2024416369}, or heterostructuring~\cite{FAN2025493, D4TA03779K}, making Quadrene a versatile building block for next-generation electronics.

\begin{figure}[ht]
    \centering
    \includegraphics[width=\textwidth]{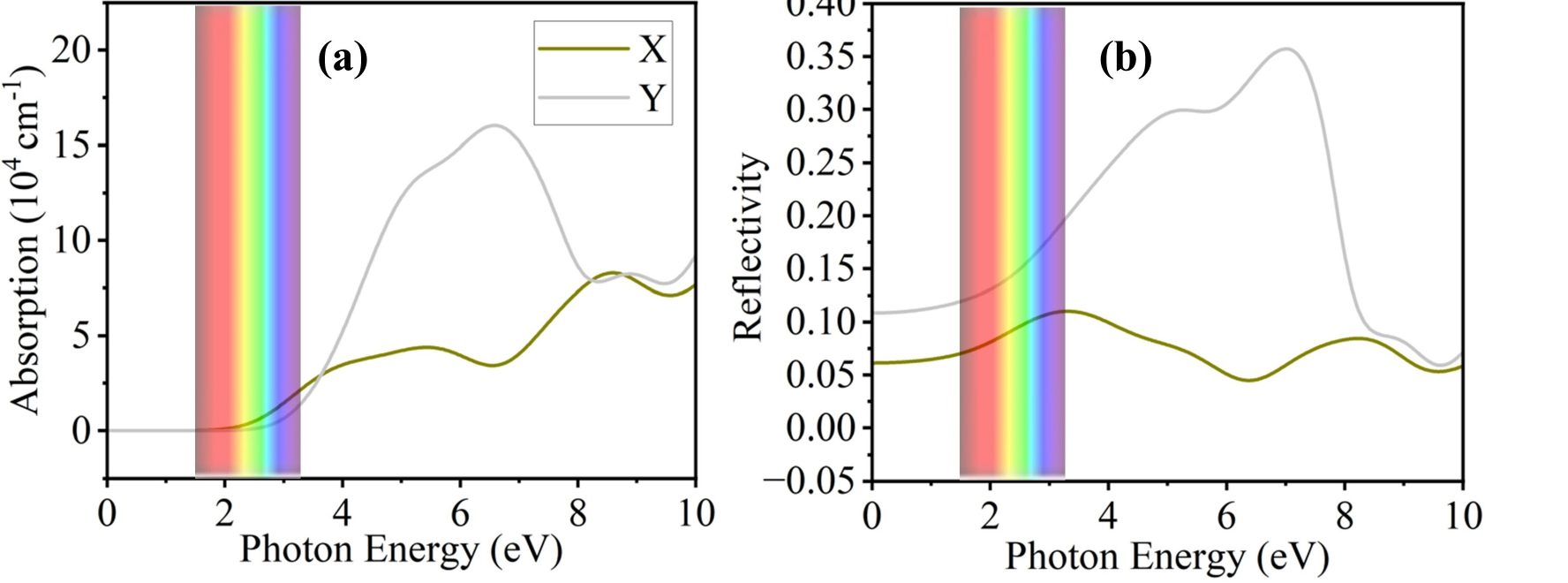}
    \caption{Optical properties of Quadrene along the X (gold) and Y (gray) in-plane directions. (a) Absorption coefficient showing stronger optical activity along the Y-direction, especially in the visible and ultraviolet range. (b) Reflectivity spectrum revealing pronounced anisotropy, with higher values along the Y-direction. The shaded region indicates the visible light spectrum.}
    \label{fig:optics}
\end{figure}

Finally, the optical response of Quadrene was investigated by calculating the frequency-dependent absorption coefficient and reflectivity, as shown in Fig.~\ref{fig:optics}. Fig.s~\ref{fig:optics}(a) and ~\ref{fig:optics}(b) present the results for the in-plane directions (X and Y), revealing pronounced optical anisotropy. 

Fig.~\ref{fig:optics}(a) shows the absorption coefficient, where the Y-direction exhibits a markedly stronger absorption profile compared to the X-direction. In particular, the Y-component reaches peak values above $1.5 \times 10^{5}$~cm$^{-1}$ in the ultraviolet region, whereas the absorption along the X-direction remains below $1.0 \times 10^{5}$~cm$^{-1}$, with modest activity even in the visible range (highlighted by the colored background). This polarization-dependent absorption indicates the potential of Quadrene for optoelectronic applications requiring directional selectivity, such as polarization-sensitive photodetectors or anisotropic solar absorbers.

Reflectivity data in Fig.~\ref{fig:optics}(b) further corroborate the optical anisotropy. The Y-polarized component shows a broad reflectance plateau with a maximum of approximately 0.36 around 5.8~eV. At the same time, the X-direction remains relatively flat and low in magnitude, not exceeding 0.12 across the entire range. The weak reflectivity in the visible regime is particularly advantageous for applications in transparent electronics and anti-reflective coatings.

\begin{figure}[ht]
    \centering
    \includegraphics[width=\textwidth]{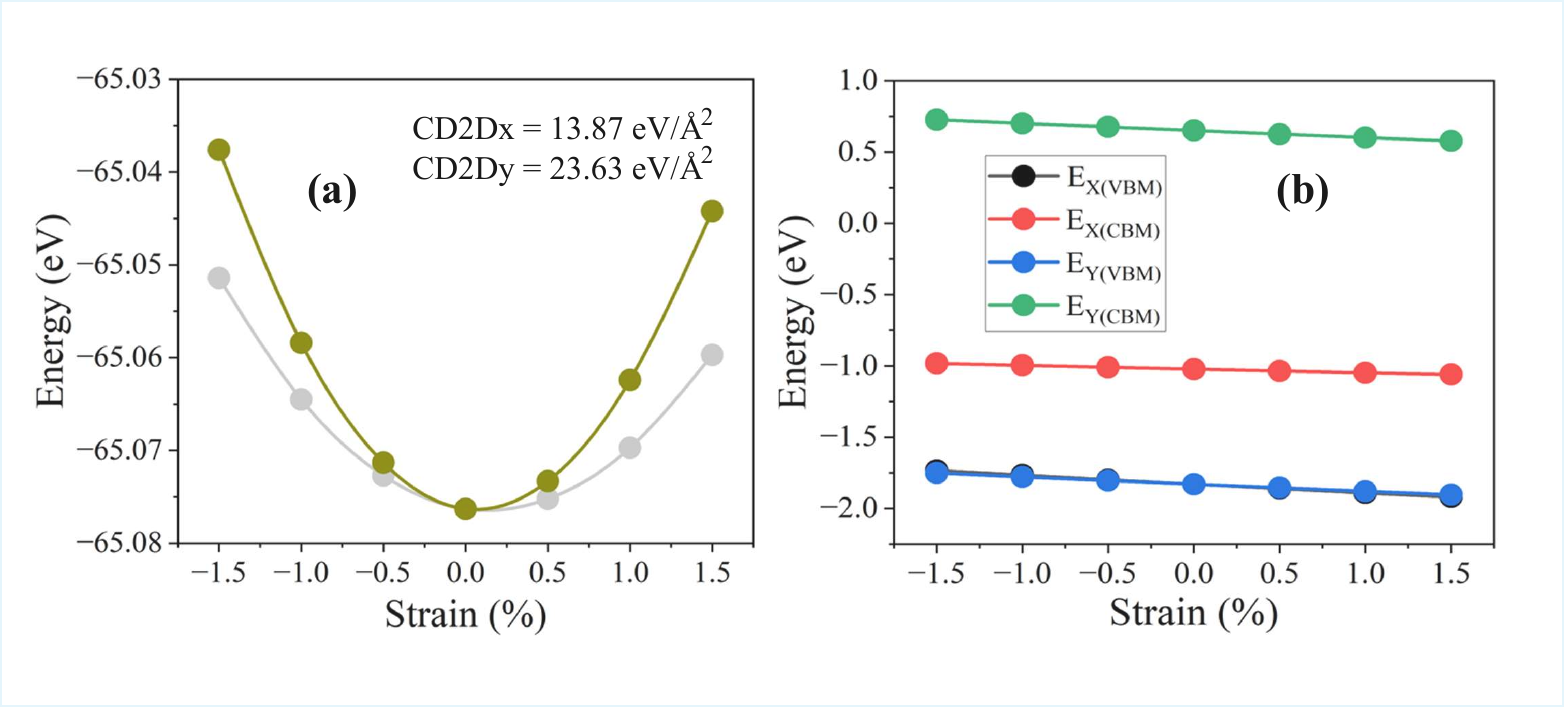}
    \caption{(a) Total energy variation as a function of uniaxial strain applied along the x and y directions, used to estimate the 2D elastic modulus (CD2D). (b) Evolution of the valence band maximum (VBM) and conduction band minimum (CBM) energies under strain, showing the strain dependence of the electronic band edges.}
    \label{fig:mobility}
\end{figure}

\begin{table*}[!htbp]
\centering
\caption{Calculated effective mass $m_i^*$, average effective mass $m_d$, in-plane stiffness $C_{2D}$, deformation potential constant $E_1$, and DP charge carrier mobility $\mu$ at $T=300$~K. Here, $m_0$ is the mass of a free electron.}
\label{mobility}
\begin{tabular}{|c|c|c|c|c|c|c|}
\hline
Direction & Carrier & $m_i^*$ ($m_0$) & $m_d$ ($m_0$) & $C_{2D}$ (eV/\AA$^2$) & $E_1$ (eV) & $\mu$ ($10^4$ cm$^2$/V$\cdot$s) \\
\hline
$x_{vbm}$ & $e$ &-0.68  & 0.62 & 13.87 & -0.061 & 200 \\
$x_{cbm}$ & $h$ & 0.83 &0.62 & 13.87 & -0.026 & 640 \\
\hline
$y_{vbm}$ & $e$ &-0.57&0.87  & 23.63 & -0.051 & 583 \\
$y_{cbm}$ & $h$ & 0.93 & 0.87 & 23.63 & -0.050 & 263 \\
\hline
\end{tabular}
\end{table*}

To evaluate the carrier mobility of Quadrene, uniaxial strains were systematically applied along both the $x$ and $y$ crystallographic directions, as shown in Figure~\ref{fig:mobility}. The strain range was set from $-1.5\%$ to $1.5\%$, ensuring a linear-elastic response and avoiding structural instabilities. For each strain configuration, the total energy and band-edge positions were computed. The total energy as a function of strain (Figure~\ref{fig:mobility}(a)) was fitted using a parabolic function, from which the in-plane elastic modulus per unit area ($C_{2D}$) was extracted. The calculated values are $C_{2D,x} = 13.87$~eV/\AA$^2$ along the $x$ direction and $C_{2D,y} = 23.63$eV/\AA$^2$ along the $y$ direction, indicating anisotropic mechanical stiffness. Concurrently, the evolution of the valence band maximum (VBM) and conduction band minimum (CBM), as well as the band gap, was linearly fitted with respect to the applied strain [Figure\ref{fig:mobility}(b)], enabling the determination of effective mass variation and deformation potential constants.

Table~\ref{mobility} summarizes the results obtained from this analysis, including the effective mass ($m^*$), deformation potential constant ($E_1$), and carrier mobility ($\mu$) for both electrons and holes. Along the $x$ direction, the hole mobility attains a value of $2.00 \times 10^{6}$~cm$^{2}$/V·s. In comparison, the electron mobility reaches $6.40 \times 10^{6}$~cm$^{2}$/V·s, attributable to the synergistic effect of a lower effective mass and reduced deformation potential. In contrast, along the $y$ direction, the hole mobility substantially increases to $5.83 \times 10^{6}$~cm$^{2}$/V·s. In contrast, the electron mobility decreases to $2.63 \times 10^{6}$~cm$^{2}$/V·s, indicating an inversion in the relative transport performance of the two carrier types. 

For comparison, TPH-I and TPH-II carbons, derived from penta-graphene through Stone–Wales transformations, exhibit anisotropic transport properties, with electron mobilities up to $\sim 7.9 \times 10^{4}$~cm$^{2}$/V·s and hole mobilities around $1.5 \times 10^{4}$~cm$^{2}$/V·s, respectively \cite{ZHANG2021147885}. Similarly, the all-sp$^3$ TTH-carbon shows carrier mobilities of about $\sim 3.0 \times 10^{3}$~cm$^{2}$/V·s along the $y$ direction, which are one to three orders of magnitude lower than Quadrene \cite{D0CP04547K}. 

Another recently proposed graphene allotrope, Me-graphene, exhibits a remarkable hole mobility of $1.6 \times 10^{5}$~cm$^{2}$/V·s \cite{D0NR03869E}. At the same time, cto-graphene, composed of triangular and 
octagonal carbon rings, demonstrates anisotropic hole mobilities of $7.3 \times 10^{4}$ and $1.3 \times 10^{4}$~cm$^{2}$/V·s along the armchair and zigzag directions, respectively \cite{doi:10.1021/acsaelm.3c00461}. These values confirm that although several carbon-based allotropes 
display promising transport performance, the carrier mobilities of Quadrene surpass them by at least one order of magnitude.

\section{Conclusion}

We have proposed and characterized a novel carbon allotrope, Quadrene, featuring a quasi-2D geometry. Its optimized lattice constants and symmetry group (\textit{Pmmm}) confirm a rectangular, anisotropic framework with a structural corrugation of 3.60~\AA{}.

Quadrene is dynamically and thermally stable, as verified by phonon dispersion and AIMD simulations at 1000~K. It exhibits marked elastic anisotropy, with Young’s moduli ranging from 65~GPa to 516~GPa and a directional Poisson's ratio up to 0.81, indicating potential for flexible nanomechanical applications.

The material exhibits semiconducting behavior with an indirect bandgap of 0.80 eV at the PBE level, which increases to 1.58 eV when calculated with the HSE06 hybrid functional. Its optical response is strongly anisotropic, with enhanced absorption and reflectivity along specific directions, making Quadrene a promising candidate for future electronic and optoelectronic devices.

Electronic transport properties show strong anisotropy. In the $x$ direction, hole and electron mobilities are $2.00 \times 10^{6}$ cm$^2$/V·s and $6.40 \times 10^{6}$ cm$^2$/V·s, respectively, thanks to low effective mass and reduced deformation potential. Along the $y$ direction, hole mobility rises to $5.83 \times 10^{6}$ cm$^2$/V·s, while electron mobility drops to $2.63 \times 10^{6}$ cm$^2$/V·s. This inversion highlights the direction-dependent transport behavior, underscoring Quadrene’s suitability for anisotropic nanoelectronic designs that enable selective directional conduction.

\section*{Acknowledgments}
This work was supported by the Brazilian funding agencies Fundação de Amparo à Pesquisa do Estado de São Paulo (FAPESP) (grants no. 2022/03959-6, 2022/14576-0, 2013/08293-7, 2020/01144-0, 2024/05087-1, and 2022/16509-9), National Council for Scientific, Technological Development (CNPq) (grants no. 307213/2021–8, 350176/2022-1, and 167745/2023-9), FAP-DF (grants no. 00193.00001808/2022-71 and 00193-00001857/2023-95), FAPDF-PRONEM (grant no. 00193.00001247/2021-20), and PDPG-FAPDF-CAPES Centro-Oeste (grant no. 00193-00000867/2024-94). 

\bibliographystyle{unsrtnat}
\bibliography{references}  

\end{document}